\documentstyle[preprint,aps,epsf]{revtex}

\newcommand{\be}{\begin{equation}}
\newcommand{\ee}{\end{equation}}

\begin{document}

\title{ Double Counting  of  `Chiral Anomaly'
      in a Model Study of Primordial Baryogenesis }

\author{ P. Ao }

\address{ Departments of Theoretical Physics, 
          Ume\aa{\ }University, S-901 87, Ume\aa, Sweden}


\maketitle
 
\begin{abstract}
{  In this comment to the Nature paper of Bevan {\it et al}
  I point out that their interpretation of experimental data is based 
  on a double counting of the `chiral anomaly' due to a 
  vortex motion: using the 
  calculation far away from the vortex core (Berry phase)
  to cancel the equivalent calculation at the core (spectral flow).
  The relaxation time approximation
  in force or momentum balance equation involved in their theory is also wrong,
  which has been rigorous proved  in transport theory since 60's. 
  Hence their affirmative conclusion is premature. }
\end{abstract}          

Arguing and exploiting the same mathematical structure in the theories for 
two totally different physical phenomena, 
Bevan {\it et al}\cite{bevan} has made the experimental study of cosmological 
baryogenesis on an Earth-bounded laboratory possible.
This new thrust of experimental studies may not only lead to new ideas in 
cosmology, also help to clear up many long standing unsolved issues 
on the mutual
frictions in superfluids of both Helium 4 and 3.\cite{donnelly}  
Here, however, I wish to point out that the theoretical basis 
for the interpretation of their data is in error, which consists of 
a double counting of the spectral flow contribution.
   
It was theoretically proposed  by Josephson\cite{josephson} and 
Anderson\cite{anderson} 
that the motion of vortices in a superfluid generates 
momentum or mass flow in a direction perpendicular to the vortex velocity.
This phase slippage mechanism of momentum generation 
has been verified experimentally, 
and is called Josephson-Anderson relation.
Acting back on the vortex, it gives rise a transverse lift force, 
the Magnus force.
Its classical manifestation is how an airplane can fly, and
is at present the only route for direct experimental study of vorticity 
quantization in both bosonic and fermionic superfluids.\cite{zieve}
For a fermionic superfluid, the Bardeen-Cooper-Schrieffer type theory
provides the connection of the transverse force to the
chiral anomaly, as has been argued by Bevan {\it et al}\cite{bevan}.
This chiral anomaly can be expressed by two completely different but 
equivalent formulae.
One form is the counting of the extended state contribution 
far away from the vortex core by the Berry phase computation\cite{at}, 
where there is no contribution from the localized core states.\cite{at,tan}
The opposite is the counting of virtual transitions, which can be solely
expressed as localized core state transitions\cite{az}, 
the statement of the spectral flow at the core.\cite{ao,az}
Thus equation (2) of Bevan {\it et al}\cite{bevan}
is an alternative formula to calculate\cite{ao}, 
not an additional transverse force assumed by them to effectively  cancel,
the Magnus force on a vortex.
Hence, their followed equations, equations (3-5), 
cannot be the consequences of the Bardeen-Cooper-Schrieffer theory.
The interpretation of the experimental data  of Bevan {\it et al}\cite{bevan} 
in this perspective
is based on counting the same force twice with opposite signs.

Practically, one may still treat equation(4) as an phenomenological proposal 
to fit experimental data\cite{hv}.
Both $d_{\perp}$ and $d_{\parallel}$ in equation (4) are simply
two fitting parameters. Their plausible justifications are supplied
subsequently.
In this way, 
certain agreement between equation (4) and the experimental data
may be obtained, as Bevan {\it et al}\cite{bevan} have found.
However, Bevan {\it et al} have also  noted that there are 
approximations in their theory, and  
one of the crucial fitting parameters, $\omega_0\tau$, 
has an activation energy whose plausible justification is absence.
One of their approximations, the relaxation time approximation in force or 
momentum balance equation, has been proved to be wrong 
in transport theory.\cite{zhu}

The conclusions may be drawn from above discussions are 
that the agreement between 
the approximated theory and a part of the experimental data is fortuitous, 
and that it is premature for Bevan {\it et al} to pronounce 
the finding of a quantitative support for a baryogenesis process.

\end{document}